\pdfoutput=1
\documentclass[prb,twocolumn,showpacs,amsmath,amssymb,floatfix,superscriptaddress]{revtex4-1}

\usepackage{graphicx}
\usepackage{dcolumn}
\usepackage{bm}
\usepackage{hyperref}

\begin{document}

\title{Annihilation of topological solitons in magnetism with spin wave burst finale: The role of nonequilibrium electrons causing nonlocal damping and spin pumping over ultrabroadband frequency range} 		
		
\author{Marko D. Petrovi\'{c}}
\affiliation{Department of Physics and Astronomy, University of Delaware, Newark, DE 19716, USA}
\author{Utkarsh Bajpai}
\affiliation{Department of Physics and Astronomy, University of Delaware, Newark, DE 19716, USA}
\author{Petr Plech\'a\v{c}}
\affiliation{Department of Mathematical Sciences, University of Delaware, Newark,  DE 19716, USA}
\author{Branislav K. Nikoli\'{c}}
\email{bnikolic@udel.edu}
\affiliation{Department of Physics and Astronomy, University of Delaware, Newark, DE 19716, USA}
		
\begin{abstract}
We not only reproduce burst of short-wavelength spin waves (SWs) observed in recent experiment [S. Woo {\em et al.}, Nat. Phys. {\bf 13}, 448 (2017)] on magnetic-field-driven annihilation of two magnetic domain walls (DWs) but, furthermore, we predict that this setup additionally generates {\em highly unusual} pumping  of electronic spin currents in the absence of any bias voltage. Prior to the instant of annihilation, their power spectrum is {\em ultrabroadband}, so they can be converted into rapidly changing in time charge currents, via the inverse spin Hall effect, as a source of THz radiation of {\em bandwidth} \mbox{$\simeq 27$ THz} where the lowest frequency is controlled by the applied magnetic field. The spin pumping stems from time-dependent fields introduced into the quantum Hamiltonian of electrons by the classical dynamics of localized magnetic moments (LMMs) comprising the domains. The pumped currents carry spin-polarized electrons which, in turn, exert {\em backaction} on LMMs in the form of nonlocal damping which is more than twice as large as conventional local Gilbert damping. The nonlocal damping can substantially modify the spectrum of emitted SWs when compared to widely-used micromagnetic simulations where conduction electrons are completely {\em absent}. Since we use fully microscopic (i.e., Hamiltonian-based)  framework, self-consistently combining time-dependent electronic nonequilibrium Green functions with the Landau-Lifshitz-Gilbert equation, we also demonstrate that previously derived phenomenological formulas miss ultrabroadband  spin pumping while  underestimating the magnitude of nonlocal damping due to {\em nonequilibrium electrons}.
\end{abstract}


\maketitle
  
\begin{figure}
	\includegraphics[scale=1.0,angle=0]{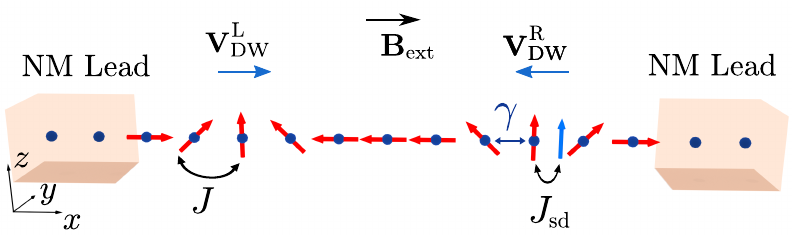}
	\caption{Schematic view of a ferromagnetic nanowire modeled as a 1D tight-binding chain whose sites host classical LMMs (red arrows) interacting with spins (blue arrow) of conduction electrons. The nanowire is attached to two NM leads terminating into the macroscopic reservoirs kept at the {\em same} chemical potential. The two DWs within the nanowire carry opposite topological charge,~\cite{Braun2012} $Q_L=-1$ for the L one and $Q_R=+1$ for the R one. They collide with the opposite velocities $\mathbf{V}^L_\mathrm{DW}$ and $\mathbf{V}^R_\mathrm{DW}$ and annihilate, upon application of an external magnetic field $\mathbf{B}_\mathrm{ext}$ parallel to the nanowire, thereby mimicking the setup of the experiment in Ref.~\onlinecite{Woo2017}.}
	\label{fig:fig1}
\end{figure}

{\em Introduction}.---The control of the domain wall (DW) motion~\cite{Tatara2008,Tatara2019,Kim2017a} within magnetic nanowires by magnetic field or current pulses is both a fundamental problem for nonequilibrium quantum many-body physics and a building block of envisaged applications in digital memories.~\cite{Parkin2015} logic~\cite{Allwood2002} and artificial neural networks.~\cite{Grollier2016} Since  DWs will be closely packed in such devices, understanding interaction between them is a problem of great interest.~\cite{Thomas2012} For example, head-to-head or tail-to-tail DWs---illustrated as the left (L) or right (R) noncollinear texture of localized magnetic moments (LMMs), respectively, in Fig.~\ref{fig:fig1}---behave as free magnetic monopoles carrying topological charge.~\cite{Braun2012} The topological charge (or the winding number) $Q \equiv -\frac{1}{\pi} \int dx\, \partial_x \phi$, associated with winding of LMMs as they interpolate between two uniform degenerate  ground states with $\phi=0$ or $\phi=\pi$, is opposite for adjacent DWs, such as $Q_L=-1$ and $Q_R=+1$ for DWs in Fig.~\ref{fig:fig1}. Thus, long-range attractive interaction between DWs can lead to their {\em annihilation}, resulting in the ground state without any DWs.~\cite{Kunz2009,Kunz2010,Ghosh2017,Kim2015} This is possible because total topological charge remains conserved, $Q_L + Q_R=0$. The {\em nonequilibrium dynamics}~\cite{Manton2004} triggered by annihilation of topological solitons is also of great interest in many other fields of physics, such as  cosmology,~\cite{Bradley2008} gravitational waves,~\cite{Nakayama2017} quantum~\cite{Manton2004} and string field~\cite{Dvali2003} theories, liquid crystals~\cite{Shen2019} and Bose-Einstein condensates.~\cite{Takeuchi2012,Nitta2012} 


The recent experiment~\cite{Woo2017} has monitored annihilation of two DWs within a metallic ferromagnetic nanowire by observing intense burst of spin waves (SWs) at the moment of annihilation.  Thus generated large-amplitude SWs are dominated by exchange, rather than dipolar, interaction between LMMs and are, therefore, of short wavelength. The SWs of $\sim 10$ nm wavelength are crucial for scalability of magnonics-based technologies,~\cite{Chumak2015,Kim2010} like signal transmission or memory-in-logic and logic-in-memory low-power digital computing  architectures. However, they are difficult to excite by other methods due to the requirement for high magnetic fields.~\cite{Navabi2017,Liu2018}

The computational simulations of DW annihilation,~\cite{Woo2017,Kunz2009,Kunz2010} together with theoretical analysis of generic features of such a phenomenon,~\cite{Ghosh2017} have been based {\em exclusively} on classical micromagnetics where one solves coupled Landau-Lifshitz-Gilbert (LLG) equations~\cite{Evans2014} for the dynamics of LMMs viewed as rotating classical vectors of fixed length.  On the other hand, the dynamics of LMMs comprising two DWs also generates time-dependent fields which will push the surrounding conduction electrons out of 
equilibrium. The {\em nonequilibrium electrons} comprise pumped spin current~\cite{Tserkovnyak2005,Petrovic2018,Chen2009} (as well as charge currents if the left-right symmetry of the device is broken~\cite{Chen2009,Bajpai2019}) in the absence of any externally applied bias voltage. The pumped spin currents flow out of the DW region into the external circuit, and since they carry away excess angular momentum of precessing LMMs, the {\em backaction} of nonequilibrium electrons on LMMs emerges~\cite{Tserkovnyak2005} as an additional damping-like (DL) spin-transfer torque (STT).  

The STT, as a phenomenon in which spin angular momentum of conduction electrons is transferred to LMMs when they are not aligned with electronic spin-polarization, is usually discussed for {\em externally injected} spin current.~\cite{Ralph2008} But here it is the result of complicated  many-body nonequilibrium state in which LMMs drive electrons out of equilibrium which, in turn, exert {\em backaction}  in the form of STT onto LMMs to modify their dynamics in a self-consistent fashion.~\cite{Petrovic2018,Sayad2015} Such effects are absent from classical  micromagnetics or atomistic spin dynamics~\cite{Evans2014} because they do not include conduction electrons. This has prompted derivation of a multitude of phenomenological expressions~\cite{Zhang2009,Kim2012,Foros2008,Tserkovnyak2009,Hankiewicz2008,Yuan2014,Yuan2016,Thonig2018} for the so-called {\em nonlocal} (i.e., magnetization-texture-dependent) and {\em spatially nonuniform} (i.e., position-dependent) Gilbert damping that could be added into the LLG equation and micromagnetics codes~\cite{Weindler2014,Wang2015,Verba2018} to capture the {\em backaction} of nonequilibrium electrons while not simulating them explicitly. Such expressions do not require spin-orbit (SO) or magnetic disorder scattering, which are necessary for conventional local Gilbert damping,~\cite{Kambersky2007,Gilmore2007,Starikov2010} but they were estimated~\cite{Kim2012,Hankiewicz2008} to be usually a small effect unless additional conditions (such as narrow DWs or intrinsic SO coupling splitting the band structure~\cite{Kim2012}) are present. On the other hand, a {\em surprising} result~\cite{Weindler2014} of Gilbert damping extracted from experiments on magnetic-field-driven DW being several times larger than the value obtained from standard ferromagnetic resonance measurements can only be accounted by including additional nonlocal damping.

\begin{figure}
	\includegraphics[scale=1.0,angle=0]{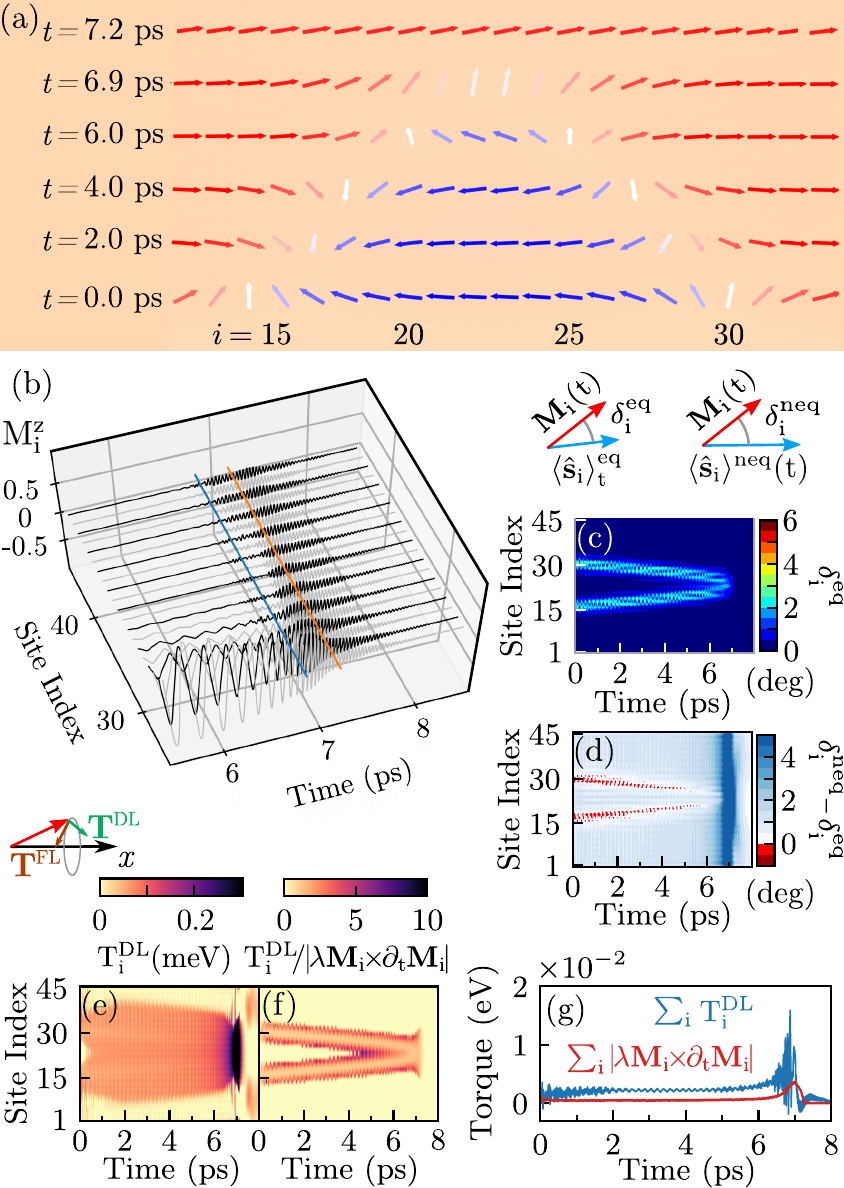}
	\caption{(a) Sequence of snapshots of two DWs, in the course of their collision and annihilation  in the setup of Fig.~\ref{fig:fig1}; and  (b) the corresponding time-dependence of the $z$-component of LMMs where blue and orange line mark \mbox{$t=6.9$ ps} (when two DWs start vanishing) and \mbox{$t=7.2$ ps} (when all LMMs become nearly parallel to the $x$-axis) from panel (a). A movie animating  panels (a) and (b) is provided in the SM.~\cite{sm} Spatio-temporal profile of: (c) angle $\delta_i^\mathrm{eq}$ and (d) ``nonadiabaticity'' angle $\delta_i^\mathrm{neq}-\delta_i^\mathrm{eq}$, with the meaning of $\delta_i^\mathrm{neq}$ and $\delta_i^\mathrm{eq}$ illustrated in the inset above panel (c); (e) DL STT  [Eq.~\eqref{eq:stt}] as electronic {\em backaction} on LMMs; (f) ratio of DL STT to conventional local Gilbert damping [Eq.~\eqref{eq:llg}]; and (g) ratio of the sum of DL STT to the sum of conventional local Gilbert damping over all LMMs.}
	\label{fig:fig2} 
\end{figure}

\begin{figure*}
	\includegraphics[width=\linewidth]{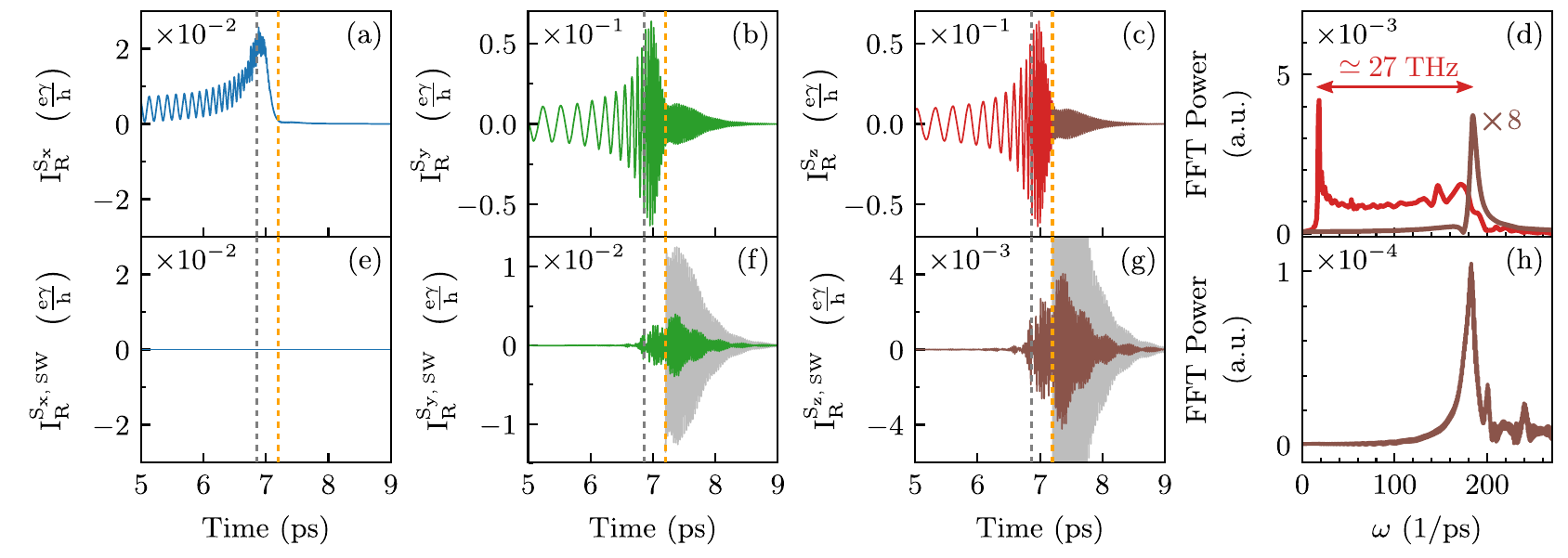}
	\caption{Time dependence of: (a)--(c) electronic spin currents  pumped into the right NM lead during DW collision and annihilation; (e)--(g) SW-generated contribution to spin currents in panels (a)--(c), respectively, after spin current carried by SW from Fig.~\ref{fig:fig2}(b) is stopped at the magnetic-nanowire/nonmagnetic-NM-lead interface and converted (as observed experimentally~\cite{Woo2017,Chumak2012})  into electronic spin current in the right NM lead. Vertical dashed lines mark times \mbox{$t=6.9$ ps} and \mbox{$t=7.2$ ps} whose snapshots of LMMs are shown in Fig.~\ref{fig:fig2}(a). For easy comparison, gray curves in panels (f) and (g) are the same as the signal in panels (b) and (c), respectively, for post-annihilation times $t \ge 7.2$ ps. Panels (d) and (h) plot FFT power spectrum of signals in panels (c) and (g), respectively, before (red curve) and after (brown curves) completed annihilation at $t = 7.2$ ps.}
	\label{fig:fig3} 
\end{figure*}

In this Letter, we unravel complicated many-body nonequilibrium state of LMMs and conduction electrons created by DW annihilation using recently developed~\cite{Petrovic2018,Bajpai2019a,Suresh2021,Suresh2020,Bostrom2019}  quantum-classical formalism which  combines time-dependent nonequilibrium Green function (TDNEGF)~\cite{Stefanucci2013,Gaury2014} description of quantum dynamics of conduction electrons with the LLG equation description of classical dynamics of LMMs on each atom.~\cite{Evans2014} Such TDNEGF+LLG formalism is fully microscopic, since it requires only the quantum Hamiltonian of electrons and the classical Hamiltonian of LMMs as input, and {\em numerically exact}.  We apply it to a  setup depicted in Fig.~\ref{fig:fig1} where two DWs reside at time $t=0$ within a  one-dimensional (1D) magnetic nanowire attached to two normal metal (NM) leads, terminating into the macroscopic reservoirs without any bias voltage. 

\begin{figure}
	\includegraphics[scale=1.0,angle=0]{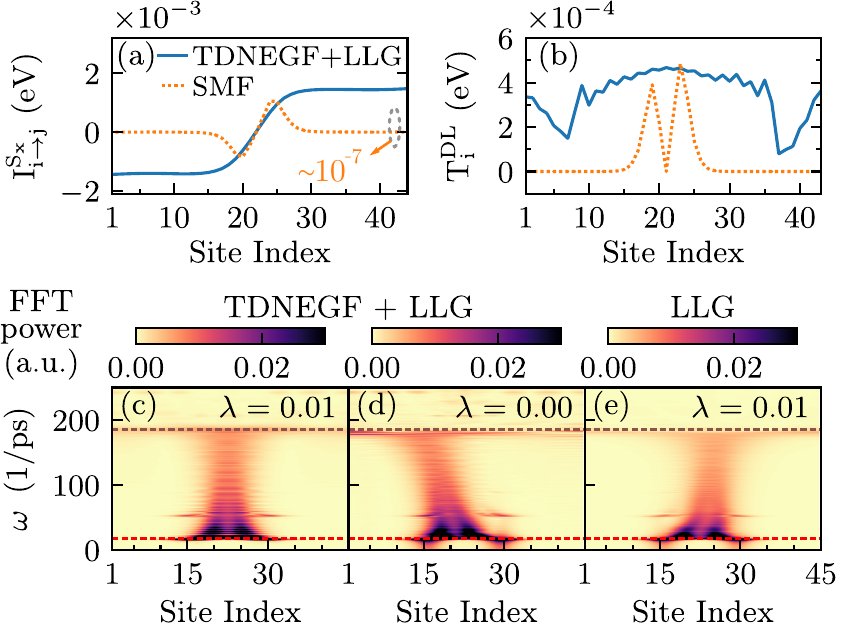}
	\caption{Spatial profile at \mbox{$t=6.9$ ps} of: (a) locally pumped spin current $I_{i \rightarrow j}^{S_x}$~\cite{Suresh2021} between sites $i$ and $j$; and nonlocal 	damping due to {\em backaction of nonequilibrium electrons}. Solid lines in (a) and (b) are obtained from TDNEGF+LLG calculations, and dashed lines are obtained from SMF theory phenomenological formulas.~\cite{Zhang2009,Kim2012,Yamane2011} (c)--(e) FFT power spectra~\cite{Kim2010} of $M_i^z(t)$ where (c) and (d) are TDNEGF+LLG-computed with $\lambda=0.01$ and $\lambda=0$, respectively, while (e) is LLG-computed with {\em backaction} of nonequilibrium electrons removed, $\mathbf{T}_i\left[\mathbf{M}_i(t)\right] \equiv 0$, in Eq.~\eqref{eq:llg}. The dashed horizontal lines in panels (c)--(e) mark frequencies of peaks in Fig.~\ref{fig:fig3}(d).}
	\label{fig:fig4} 
\end{figure}

Our {\em principal  results} are: ({\em i}) annihilation of two DWs [Fig.~\ref{fig:fig2}] pumps highly unusual electronic spin currents whose power spectrum is {\em ultrabroadband} prior to the instant of annihilation [Fig.~\ref{fig:fig3}(d)], unlike the narrow peak around a single frequency for standard spin pumping;~\cite{Tserkovnyak2005} ({\em ii}) because pumped spin currents carry  away excess angular momentum of precessing LMMs, this acts as DL STT on LMMs which is spatially [Figs.~\ref{fig:fig2}(e) and ~\ref{fig:fig4}(b)] and time [Fig.~\ref{fig:fig2}(g)] dependent, as well as \mbox{$\simeq 2.4$} times {\em larger} [Fig.~\ref{fig:fig2}(f)] than conventional local Gilbert damping [Eq.~\eqref{eq:llg}]. This turns out to be {\em remarkably similar} to \mbox{$\simeq 2.3$} ratio of nonlocal and local Gilbert damping measured experimentally in permalloy,~\cite{Weindler2014} but it is severely underestimated by phenomenological theories~\cite{Zhang2009,Kim2012} [Fig.~\ref{fig:fig4}(a),(b)].

{\em Models and methods}.---The classical Hamiltonian for 
LMMs, described by unit vectors $\mathbf{M}_i(t)$ at each site $i$ of 1D lattice, is chosen as
\begin{eqnarray}
\label{eq:classH}
{\mathcal H} & = & -J \sum_{\langle i j \rangle}
\mathbf{M}_{i}\cdot \mathbf{M}_{j}
- K\sum_{i}
{\left(M_{i}^{x}\right)}^2 \nonumber \\
& &
+ D\sum_{i}
{\left(M_{i}^{y}\right)}^2 
-\mu_{B} \sum_{i} \mathbf{M}_i\cdot
\mathbf{B}_{\rm ext},
\end{eqnarray}
where \mbox{$J = 0.1\ {\rm eV}$} is the Heisenberg exchange coupling between the nearest-neighbor LMMs; $K = 0.05\ {\rm eV}$ is the magnetic anisotropy along the $x$-axis; and $D = 0.007\ {\rm eV}$ is the demagnetizing field along the $y$-axis. The last term in Eq.~\eqref{eq:classH} is the Zeeman energy ($\mu_{B}$ is the Bohr magneton) describing the interaction of LMMs with an external magnetic field $\mathbf{B}_{\rm ext}$ parallel to the nanowire in Fig.~\ref{fig:fig1} driving the DW dynamics, as employed in the experiment.~\cite{Woo2017}  The classical dynamics of LMMs is described by a system of coupled LLG equations~\cite{Evans2014} (using notation $\partial_t \equiv \partial/\partial t$) 
\begin{eqnarray}\label{eq:llg}
\partial_t \mathbf{M}_i & = & -g \mathbf{M}_i \times \mathbf{B}^\mathrm{eff}_i + \lambda \mathbf{M}_i \times \partial_t \mathbf{M}_i \nonumber \\
&& + \frac{g}{\mu_M}\left( \mathbf{T}_i \left[I^{S_\alpha}_\mathrm{ext}\right] + \mathbf{T}_i\left[\mathbf{M}_i(t)\right] \right).  
\end{eqnarray}
where \mbox{$\mathbf{B}^{\rm eff}_i = - \frac{1}{\mu_M} \partial \mathcal{H} /\partial \mathbf{M}_{i}$}  is the effective magnetic field ($\mu_M$ is the magnitude of LMMs); $g$ is the gyromagnetic ratio; and the magnitude of conventional local Gilbert damping  is specified by spatially- and time-independent $\lambda$, set as $\lambda=0.01$ as the typical value measured~\cite{Weindler2014} in metallic ferromagnets. The spatial profile of a single DW in equilibrium, i.e., at time $t = 0$ as the initial condition, is given by \mbox{${\bf M}_i(Q, X_{\rm DW}) = 
	\big(\cos{\phi_i(Q, X_{\rm DW})},
	\ 0,
	\ \sin{\phi_i(Q, X_{\rm DW})}\big)$}, 
where ${\phi_i}(Q, X_{\rm DW}) = Q\arccos\left[\tanh\left(x_i - X_{\rm DW}\right)\right]$; $Q$ is the topological charge; and $X_{\rm DW}$ is the
position of the DW. The initial configuration of two DWs, $\mathbf{M}_{i}(t = 0) =  \mathbf{M}_{i}(Q_{\rm L}, X_{\rm L}) +
	\mathbf{M}_{i}(Q_{\rm R}, X_{\rm R})$, positioned at sites $X_{\rm L} = 15$ and  $X_{\rm R} = 30$ harbors opposite topological charges  $Q_{\rm R} = -Q_{\rm L} = 1$ around these sites. 

In general, two additional terms~\cite{Zhang2004,Zhang2009,Kim2012} in Eq.~\eqref{eq:llg} extend the original LLG equation---STT due to externally injected electronic spin current,~\cite{Ralph2008} which is actually {\em absent}  $\mathbf{T}_i \left[I^{S_\alpha}_\mathrm{ext}\right] \equiv 0$  in the setup of Fig.~\ref{fig:fig1}; and STT due to {\em backaction} of electrons
\begin{equation}\label{eq:stt}
\mathbf{T}_i\left[\mathbf{M}_i(t) \right]=J_\mathrm{sd} (\langle \hat{\mathbf{s}}_i\rangle^\mathrm{neq}(t)- \langle \hat{\mathbf{s}}_i \rangle^\mathrm{eq}_t) \times \mathbf{M}_i(t),
\end{equation} 
driven out of equilibrium by $\mathbf{M}_i(t)$. Here  \mbox{$J_{\rm sd}=0.1$ eV} is chosen as the $s$-$d$ exchange coupling (as measured in permalloy~\cite{Cooper1967}) between LMMs and electron spin. We obtain ``adiabatic''~\cite{Stahl2017,Bajpai2020} electronic spin density, \mbox{$\langle \hat{\mathbf{s}}_i\rangle^\mathrm{eq}_t=\mathrm{Tr} \, [{\bm \rho}^\mathrm{eq}_t|i\rangle \langle i| \otimes {\bm \sigma}]$}, from grand canonical equilibrium density matrix (DM) for instantaneous configuration of $\mathbf{M}_i(t)$ at time $t$ [see Eq.~\eqref{eq:eqrho}]. Here \mbox{${\bm  \sigma} = (\hat{\sigma}_x,\hat{\sigma}_y,\hat{\sigma}_z)$} is the vector of the Pauli matrices. The nonequilibrium electronic spin density, \mbox{$\langle \hat{\mathbf{s}}_i \rangle^\mathrm{neq}(t) = \mathrm{Tr} \, [{\bm \rho}_\mathrm{neq}(t) |i\rangle \langle i| \otimes {\bm \sigma}]$}, requires to compute  time-dependent nonequilibrium DM, \mbox{${\bm \rho}_\mathrm{neq}(t)= \hbar \mathbf{G}^<(t,t)/i$}, which we construct using TDNEGF algorithms explained in Refs.~\onlinecite{Croy2009,Popescu2016} and combine~\cite{Petrovic2018} with the classical LLG equations [Eq.~\eqref{eq:llg}] using time step \mbox{$\delta t=0.1$ fs}. The TDNEGF calculations require as an input a quantum Hamiltonian for electrons, which is chosen as the tight-binding one 
\begin{equation}\label{eq:tbh}
\hat{H}(t) = -\gamma \sum_{\langle ij \rangle}  \hat{c}_{i}^\dagger\hat{c}_i - J_\mathrm{sd} \sum_{i}\hat{c}_i^\dagger\boldsymbol{\sigma} \cdot \bold{M}_i(t) \hat{c}_i.
\end{equation} 
Here \mbox{$\hat{c}_i^\dagger = (\hat{c}_{i\uparrow}^\dagger,\hat{c}_{i\downarrow}^\dagger)$}  is a row vector containing operators $\hat{c}_{i\sigma}^\dagger$ which create an electron of spin $\sigma=\uparrow,\downarrow$ at the site $i$, and $\hat{c}_i$ is a column vector that contains the corresponding  annihilation operators; and \mbox{$\gamma=1$ eV}  is the nearest-neighbor hopping. The magnetic nanowire in the setup in Fig.~\ref{fig:fig1} consists of 45 sites and it is attached to semi-infinite NM leads modeled by the first term  in  $\hat{H}$. The Fermi energy of the reservoirs is set at  \mbox{$E_F=0$~eV}. Due to the  computational complexity of TDNEGF calculations,~\cite{Gaury2014} we use magnetic field  $|\mathbf{B}_{\rm ext}| = 100 \ {\rm T}$  to complete DW annihilation on $\sim$ ps time scale (in the experiment~\cite{Woo2017} this happens within \mbox{$\sim 2$ ns}). 

{\em Results}.---Figure~\ref{fig:fig2}(a) demonstrates that TDNEGF+LLG-computed snapshots of $\mathbf{M}_i(t)$ {\em fully reproduce} annihilation 
in the experiment,~\cite{Woo2017} including {\em finale} when SW burst is emitted at $t \simeq 7.2$ ps in Fig.~\ref{fig:fig2}(b). The corresponding complete spatio-temporal profiles are animated as a movie provided in the Supplemental Material (SM).~\cite{sm} However, in contrast to micromagnetic simulations of Ref.~\onlinecite{Woo2017} where electrons are absent, Fig.~\ref{fig:fig2}(d) shows that they generate spin density $\langle \hat{\mathbf{s}}_i\rangle^\mathrm{neq}(t)$ which is {\em noncollinear} with either  $\mathbf{M}_i(t)$ or $\langle \hat{\mathbf{s}}_i \rangle^\mathrm{eq}_t$. This leads to ``nonadiabaticity'' angle $(\delta_i^\mathrm{neq} - \delta_i^\mathrm{eq}) \neq  0$ in Fig.~\ref{fig:fig2}(d) and nonzero STT [Eq.~\eqref{eq:stt} and Fig.~\ref{fig:fig2}(e)] as self-consistent {\em backaction} of conduction electrons onto LMMs driven out of equilibrium by the dynamics of LMMs themselves. The STT vector, \mbox{$\mathbf{T}_i=\mathbf{T}^\mathrm{FL}_i + \mathbf{T}^\mathrm{DL}_i$}, can be decomposed [see inset above Fig.~\ref{fig:fig2}(e)] into: ({\em i}) even under  time-reversal or field-like (FL) torque,  which affects precession of LMM around $\mathbf{B}^{\rm eff}_i$; and  ({\em ii}) odd under time-reversal or DL torque, which either enhances Gilbert term [Eq.~\eqref{eq:llg}] by pushing LMM toward $\mathbf{B}^{\rm eff}_i$ or competes with it  as antidamping. Figure~\ref{fig:fig2}(f) shows that  $\mathbf{T}_i^\mathrm{DL} \left[\mathbf{M}_i(t) \right]$ acts like an additional nonlocal damping while being $\simeq 2.4$ times larger than conventional local Gilbert damping $\lambda \mathbf{M}_i \times \partial_t \mathbf{M}_i$ [Eq.~\eqref{eq:llg}].

The quantum transport signature of DW vanishing within the time interval \mbox{$t = 6.9$--$7.2$ ps} in Fig.~\ref{fig:fig2}(a) is the reduction in the magnitude of pumped electronic spin currents [Fig.~\ref{fig:fig3}(a)--(c)]. In fact, $I^{S_x}_R(t) \rightarrow 0$  becomes zero [Fig.~\ref{fig:fig3}(a)] at \mbox{$t = 7.2$ ps} at which LMMs in Fig.~\ref{fig:fig2}(a) turn nearly parallel to the $x$-axis while precessing around it. The frequency power spectrum [red curve in Fig.~\ref{fig:fig3}(d)] obtained from fast Fourier transform (FFT) of $I^{S_z}_\mathrm{R}(t)$, for times prior to completed annihilation and SW burst at $t = 7.2$ ps, reveal highly unusual spin pumping over an {\em ultrabroadband} frequency range. This can be contrasted with the usual spin  pumping~\cite{Tserkovnyak2005} whose power spectrum is just a peak around a single frequency,~\cite{Bocklage2017} as also obtained [brown curve in Fig.~\ref{fig:fig3}(d)] by FFT of $I^{S_z}_R(t)$ at post-annihilation times  $t > 7.2$ ps.

The spin current in Fig.~\ref{fig:fig3}(a)--(c) has contributions from both electrons moved by time-dependent $\mathbf{M}_i(t)$ and SW hitting the magnetic-nanowire/NM-lead interface. At this interface, SW spin current is stopped and transmuted~\cite{Bauer2011,Suresh2021,Suresh2020} into an electronic spin current flowing into the NM lead. The transmutation is often employed  experimentally for direct electrical detection of SWs, where an electronic spin current on the NM side is converted into a voltage signal via the inverse spin Hall effect.~\cite{Woo2017,Chumak2012} Within the TDNEGF+LLG picture, SW reaching the last LMM of the magnetic nanowire, at the sites $i=1$ or $i=45$ in our setup, initiates their dynamics whose coupling to conduction electrons in the neighboring left and right NM leads, respectively, leads to pumping~\cite{Suresh2021} of the electronic spin current into the NM leads. The   properly isolated electronic spin current due to transmutation of SW burst, which we denote by $I^{S_\alpha,\mathrm{SW}}_p$,  is either zero or very small until the burst is generated in Fig.~\ref{fig:fig3}(e)--(g), as expected. We note that  detected spin current in the NM leads was attributed in the experiment~\cite{Woo2017} solely to SWs, which corresponds in our picture to considering only $I^{S_\alpha,\mathrm{SW}}_p$ while disregarding $I^{S_\alpha}_p - I^{S_\alpha,\mathrm{SW}}_p$.

{\em Discussion}.---A computationally simpler alternative to our numerical self-consistent TDNEGF+LLG is to ``integrate out  electrons''~\cite{Sayad2015,Onoda2006,Nunez2008,Fransson2008,Hurst2020} and derive effective 
expressions solely in terms of $\mathbf{M}_i(t)$, which can then be added into the LLG Eq.~\eqref{eq:llg} and micromagnetics codes.~\cite{Weindler2014,Wang2015,Verba2018} For example, spin motive force (SMF) theory~\cite{Yamane2011} gives $I_\mathrm{SMF}^{S_x}(x)=\frac{g\mu_B \hbar G_0}{4e^2} [ \partial {\bf M}(x, t)/\partial t \times \partial {\bf M}(x, t)/\partial x ]_x$ for the spin current pumped by dynamical magnetic texture, so that \mbox{$\mathbf{M} \times \mathcal{D} \cdot \partial_t \mathbf{M}$} is the corresponding  nonlocal Gilbert damping.~\cite{Zhang2009,Kim2012} Here $\mathbf{M}(x,t)$ is local magnetization (assuming our 1D system); $\mathcal{D}_{\alpha\beta}=\eta \sum_\nu (\mathbf{M} \times \partial_\nu \mathbf{M})_\alpha (\mathbf{M} \times \partial_\nu \mathbf{M})_\beta$ (using  notation $\alpha,\beta,\nu \in \{x,y,z\}$) is \mbox{$3 \times 3$} spatially-dependent damping tensor; and \mbox{$\eta=\frac{g\mu_B \hbar G_0}{4e^2}$} with $G_0 = G^\uparrow + G^\downarrow$ being the total conductivity. We compare in Fig.~\ref{fig:fig4}: ({\em i}) spatial profile of $I_\mathrm{SMF}^{S_x}(x)$ to locally pumped spin current $I_{i \rightarrow j}^{S_x}$~\cite{Suresh2021} from TDNEGF+LLG calculations [Fig.~\ref{fig:fig4}(a)] to find that the former predicts negligible spin current flowing into the leads, thereby missing {\em ultrabroadband} spin pumping predicted in Fig.~\ref{fig:fig3}(d); ({\em ii}) spatial profile of \mbox{$\mathbf{M} \times \mathcal{D} \cdot \partial_t \mathbf{M}$} to DL STT $T_i^\mathrm{DL}$ from TDNEGF+LLG calculations, to find that the former has comparable magnitude only within the DW region but with substantially differing profiles. Note also that~\cite{Suresh2021} $\left[ \sum_{i} \bold{T}_i(t) \right]_\alpha = \frac{\hbar}{2e} \left[ I^{S_\alpha}_L(t) + I^{S_\alpha}_R(t) \right] +  \sum_{i} \frac{\hbar}{2} \frac{\partial\langle\hat{\mathrm{s}}^\alpha_i\rangle^\mathrm{neq}}{\partial t}$, which makes the sum of DL STT plotted in Fig.~\ref{fig:fig2}(g) time-dependent during collision, in contrast to the sum of local Gilbert damping shown in Fig.~\ref{fig:fig2}(g). The {\em backaction of nonequilibrium electrons} via $\mathbf{T}_i \left[\mathbf{M}_i(t) \right]$  can strongly affect the dynamics of LMMs, especially for the case of short wavelength SWs and narrow DWs,~\cite{Zhang2009,Kim2012,Wang2015,Verba2018} as confirmed by comparing FFT power spectra of $M_i^z(t)$ computed by TDNEGF+LLG [Fig.~\ref{fig:fig4}(c),(d)] with those from LLG calculations [Fig.~\ref{fig:fig4}(e)] without any {\em backaction}.

We note that SMF theory~\cite{Yamane2011} is derived in the ``adiabatic'' limit,~\cite{Tatara2019,Stahl2017} which assumes that electron spin remains in the the lowest energy state at each time. ``Adiabaticity'' is used in two different contexts in spintronics with noncollinear magnetic textures---temporal and spatial.~\cite{Tatara2019} In the former case, such as when electrons interact with classical macrospin due to collinear LMMs, one assumes 
that classical spins are slow and $\langle \hat{\mathbf{s}}_i\rangle^\mathrm{neq}(t)$ can ``perfectly lock''~\cite{Tatara2019} to the direction $\mathbf{M}_i(t)$ of LMMs. In the latter case, such as for electrons traversing thick DW, one assumes that electron spin keeps the lowest energy state by rotating according to the orientation of $\mathbf{M}_i(t)$ at each spatial point, thereby evading reflection from the texture.~\cite{Tatara2019}  The concept of ``adiabatic'' limit is made a bit more quantitative by considering~\cite{Tatara2019} ratio of relevant energy scales,  $J_\mathrm{sd}/\hbar \omega \gg 1$ or  $J_\mathrm{sd}/\mu_B |\mathbf{B}_{\rm ext}| \gg 1$, in the former case; or combination of energy and spatial scales, $J_\mathrm{sd} d_\mathrm{DW}/\hbar v_F = J_\mathrm{sd} d_\mathrm{DW}/\gamma a \gg 1$, in the latter case (where $v_F$ is the Fermi velocity, $a$ is the lattice spacing and  $d_\mathrm{DW}$ is the DW thickness). In our simulations, \mbox{$J_\mathrm{sd}/\mu_B |\mathbf{B}_{\rm ext}| \approx 10$} and \mbox{$J_\mathrm{sd} d_\mathrm{DW}/\gamma a \approx 1$} for $d_\mathrm{DW} \approx 10a$ in Fig.~\ref{fig:fig2}(a). Thus,  it seems that fair comparison of our results to SMF theory requires to substantially increase $J_\mathrm{sd}$. However, \mbox{$J_\mathrm{sd}=0.1$ eV} (i.e., $\gamma/J_\mathrm{sd}/ \sim 10$, for typical $\gamma \sim 1$ eV which controls how fast is quantum dynamics of electrons) in our simulations is fixed by measured properties of permalloy.~\cite{Cooper1967}

Let us recall that rigorous definition of ``adiabaticity'' assumes that conduction electrons within a closed quantum system~\cite{Stahl2017} at time $t$ are in the ground state $|\Psi_0\rangle$ for the given configuration of LMMs $\mathbf{M}_i(t)$, $|\Psi(t)\rangle = |\Psi_0[\mathbf{M}_i(t)]\rangle$; or in open  system~\cite{Bajpai2020} their quantum state is specified by grand canonical DM
\begin{equation}\label{eq:eqrho}
	{\bm \rho}^\mathrm{eq}_t = -\frac{1}{\pi} \int dE\, \mathrm{Im} \mathbf{G}^r_t f(E).
\end{equation}
where the retarded GF, $\mathbf{G}^r_t=\big[E - \mathbf{H}[\mathbf{M}_i(t)] - {\bm \Sigma}_L - {\bm \Sigma}_R \big]^{-1}$, and ${\bm \rho}^\mathrm{eq}_t$ depend parametrically~\cite{Bode2011,Thomas2012a,Mahfouzi2016} (or implicitly, so we put $t$ in the subscript) on time via instantaneous configuration of $\mathbf{M}_i(t)$, thereby effectively assuming \mbox{$\partial_t  \mathbf{M}_i(t)= 0$}. Here $\mathrm{Im} \mathbf{G}^r_t=(\mathbf{G}^r_t - [\mathbf{G}^r_t]^\dagger)/2i$; ${\bm \Sigma}_{L,R}$ are self-energies due to the leads; and $f(E)$ is the Fermi function. For example, the analysis of Ref.~\onlinecite{Yamane2011} assumes $\langle \hat{\mathbf{s}}_i\rangle^\mathrm{neq}(t) \parallel \langle \hat{\mathbf{s}}_i\rangle^\mathrm{eq}_t$ to reveal the origin of spin and charge pumping in SMF theory---nonzero angle $\delta_i^\mathrm{eq}$ between $\langle  \hat{\mathbf{s}}_i\rangle^\mathrm{eq}_t$ and $\mathbf{M}_i(t)$ with the transverse component scaling $|\langle \hat{\mathbf{s}}_i\rangle^\mathrm{eq}_t  \times \mathbf{M}_i(t)| / \big(\langle \hat{\mathbf{s}}_i\rangle^\mathrm{eq}_t \cdot  \mathbf{M}_i(t) \big) \propto 1/J_\mathrm{sd}$ as the signature of ``adiabatic'' limit. Note that our $\delta_i^\mathrm{eq} \lesssim 4^\circ$ [Fig.~\ref{fig:fig2}(c)] in the region of two DWs (and $\delta_i^\mathrm{eq} \rightarrow 0$ elsewhere). Additional Figs.~S1--S3 in the SM,~\cite{sm} where we isolate two neighboring LMMs from the right DW in Fig.~\ref{fig:fig1} and put them in steady precession with frequency $\omega$ for simplicity of analysis, demonstrate that entering such ``adiabatic'' limit requires unrealistically large \mbox{$J_\mathrm{sd} \gtrsim 2$ eV}. Also, our exact~\cite{Bajpai2020} result [Figs.~S1(b), S2(b) and S3(b) in the SM~\cite{sm}] shows  $|\langle \hat{\mathbf{s}}_i\rangle^\mathrm{eq}_t \times \mathbf{M}_i(t)| / \big(\langle \hat{\mathbf{s}}_i\rangle^\mathrm{eq}_t \cdot  \mathbf{M}_i(t) \big) \propto 1/J_\mathrm{sd}^2$ (instead of  $\propto  1/J_\mathrm{sd}$ of Ref.~\onlinecite{Yamane2011}). Changing $\hbar \omega$---which, according to Fig.~\ref{fig:fig3}(c), is effectively increased by the dynamics of annihilation from $\hbar \omega \simeq 0.01$ eV, set initially by  $\mathbf{B}_{\rm ext}$,  toward  $\hbar \omega \simeq 0.1$ eV---only modifies  scaling of the transverse component of $\langle \hat{\mathbf{s}}_i\rangle^\mathrm{neq}(t)$ with $J_\mathrm{sd}$ [Figs.~S1(a), ~S2(a), S3(a), S4(b) and S4(d) in the SM~\cite{sm}].  The nonadiabatic corrections~\cite{Bajpai2020,Bode2011,Thomas2012a,Mahfouzi2016} take into account \mbox{$\partial_t  \mathbf{M}_i(t) \neq 0$}. We note that only in the limit \mbox{$J_\mathrm{sd} \rightarrow \infty$}, $\big( \langle \hat{\mathbf{s}}_i\rangle^\mathrm{neq}(t) - \langle \hat{\mathbf{s}}_i\rangle^\mathrm{eq}_t \big) \rightarrow 0$. Nevertheless, multiplication of these two limits within Eq.~\eqref{eq:stt} yields nonzero geometric STT,~\cite{Stahl2017,Bajpai2020} as signified by $J_\mathrm{sd}$-independent  STT [Figs.~S1(c), S2(c) and S3(c) in the SM~\cite{sm}]. Otherwise, ``nonadiabaticity'' angle is always present  $(\delta_i^\mathrm{neq} - \delta_i^\mathrm{eq}) \neq  0$ [Fig.~\ref{fig:fig2}(d)], even in the absence of spin relaxation due to magnetic impurities or SO coupling,~\cite{Evelt2017} and it can be directly related to additional spin and charge pumping~\cite{Suresh2020,Evelt2017} [see also Figs.~S1(f), S2(f) and S3(f) in the SM~\cite{sm}].

{\em Conclusions and outlook}.---The pumped spin current over {\em ultrabroadband} frequency range [Fig.~\ref{fig:fig3}(d)], as our central prediction,  can be converted into rapidly changing transient charge current via the inverse spin Hall effect.~\cite{Wei2014,Seifert2016,Chen2019} Such charge current will, in turn, emit electromagnetic radiation covering $\sim 0.03$--$27$ THz range (for $|\mathbf{B}_\mathrm{ext}| \sim 1$ T) or $\sim 0.3$--$27.3$ THz range (for $|\mathbf{B}_\mathrm{ext}| \sim 10$ T), which is highly sought range of frequencies for variety of applications.~\cite{Seifert2016,Chen2019}

\begin{acknowledgments}
M.~D.~P., U.~B., and B. K. N.  was supported by the US National Science Foundation (NSF) Grant No. ECCS 1922689. P.~P. was supported by the US Army Research Office (ARO) MURI Award No. W911NF-14-0247. 
\end{acknowledgments}


\end{document}